\documentstyle[prb,aps,multicol,graphicx,psfrag]{revtex}

\begin{document} 
\draft

\onecolumn

\title{Tuning of dynamic localization in coupled mini-bands:  signatures
of a field induced insulator-metal transition}

\author{P.  H.  Rivera$^1$ and P.  A.  Schulz$^2$}

\address{$^1$ Departamento de F\'{\i}sica, Universidade Federal de S\~ao
Carlos \\ 13564--200, S\~ao Carlos, SP, Brazil}

\address{$^2$ Instituto de F\'{\i}sica 'Gleb Wataghin', Universidade
Estadual de Campinas \\ 13083-970 Campinas, S\~ ao Paulo, Brazil}

\date{\today}

\maketitle 
\begin{abstract} 
We follow the evolution of dressed coupled
mini-bands as a function of an AC field intensity, non perturbatively, for
a wide field frequency range.  High and low frequency limits are
characterized by two different dynamic localization regimes, clearly
separated by a breakdown region in a quasi-energy map.  Signatures of a
insulator-metal like transition by means of a field induced suppression of
a Peierls-like instability are identified.  
\end{abstract}


\begin{multicols}{2}

The behaviour of semiconductor micro-structures driven by intense AC
fields, has been of renewed interest, from both theoretical and
experimental points of view, in the past few years
\cite{dunlap,holt92,wag94,pawel95,jauho95,plat97,paulo,keay,ooster}. This
subject, originally addressed in the context of atomic and molecular
physics \cite{chu85}, raised the question of how these effects would
appear in a periodic lattice, where the atomic levels are broadened into
bands \cite{brandi}.  Among the theoretical predictions, we pay special
attention on a pioneering one \cite{holt92}:  the collapses of isolated
mini-bands, leading to dynamic localization in finite superlattices (SLs).
The period of these SLs are typically orders of magnitude
shorter than the radiation wavelengths for energy scales mentioned above.
Consequently, the dipole approximation holds and suggests a common
framework for superlattices driven by intense FIR lasers or by high
frequency bias.

The problem of multi-mini-bands has also been initially addressed few years
ago by Holthaus and Hone \cite{holt94}. The
same authors suggest an interesting possibility in reducing the multi-band
problem to one of two almost symmetric bands, by means of SL dimerization
\cite{hone93}.  By alternating quantum wells (QWs) of two different widths
one could design a system where two almost symmetric mini-bands may be
strongly coupled by a field, whereas isolated from other electronic
states.  A second procedure is based on alternating barrier widths. The
quasi-energy spectra as a function of field intensity for both
dimerization procedures have been investigated theoretically in the past
few years \cite{zhao97,bao98}.  More general models \cite{jauho95,drese96}
combine characteristics of both procedures and show a quite
complex behaviour and different localization regimes are not clearly
separated, as will be discussed.

In this letter we show that coupling between mini-bands lead to new and
unexpected phenomena:  double mini-bands of dimerized SLs (alternating
well widths) may present two different dynamic localization mechanisms as
a function of field intensity.  These mechanisms are clearly separated by
a dynamical breakdown region in a map of the mini-band width on a mini-band
gap/field intensity representation space.  Besides, for resonant
field frequencies, signatures of a field induced phase transition are
observed:  at certain field intensities the effective dimerization is
suppressed and an analog to an insulator-metal transition takes place.

 We focus on a heuristic two-mini-band model in the presence of a strong AC
electric field, within a tight-binding framework, emulating the dimerized
SL proposal \cite{hone93}.  A linear chain is considered for this SL,
where each single ``atomic" $s$-like orbital of the site is associated to
one quantized level energy of a QW.  The hopping parameters describe the
coupling between the QW levels through the SL barriers.  The applied AC
fields are parallel to the chain.  Hence, our model is described by the
Hamiltonian $ H= H_o+H_{int}$, considering nearest neighbour interaction 
only:

$$ 
H_o=\sum_{\ell}\epsilon_{\ell}|\ell><\ell|+
$$
\begin{equation}
{V\over2}\sum_{\ell}\Bigl[|\ell><\ell+1|+|\ell+1><\ell|\Bigr] 
\end{equation}
\begin{equation} 
H_{int}=eaF\cos \omega t\sum _{\ell}|\ell>\ell<\ell|
\end{equation}

\noindent 
where $\epsilon_{\ell}=(-1)^{\ell}E_g/2$, $E_g$ is the mini-band gap 
energy at zero field and $\ell $ is the index site; 
$\omega ,\ F$ are the AC field frequency and amplitude,
respectively.  $a$ is the SL period, $e$ is the electron charge.  For
$E_g=0$ one has the single band limit, while for $E_g\neq 0$ one 
has a dimerized SL with a $2a$ period.  The treatment of the
time-dependent problem is based on Floquet states $|\ell,m>$ where $m$ is
the photon index.  We follow the procedure first developed by Shirley
\cite{shirley} which consists in the transformation of the time-dependent
Hamiltonian into a time-independent infinite matrix which must be
truncated.  The matrix elements are:

$$
[({\mathcal E}-m\hbar\omega -\epsilon_{\ell})\delta_{\ell '\ell}-{V\over2}(\delta_{\ell',\ell-1}+\delta_{\ell',\ell+1})]\delta _{m'm}= 
$$ 
\begin{equation} 
F'\ell\delta_{\ell'\ell}(\delta _{m',m-1}+\delta _{m',m+1}) 
\end{equation}

\noindent 
where $F'={1\over2}eaF$.  The dimension of the matrix is
$L(2M+1)$, where $L$ is the number of atomic sites, while $M$ is the
maximum photon index.  We choose $M$ in order to satisfy a convergence
condition:  symmetric spectra relative to the Quasi Brillouin Zones (QBZs)
edges\cite{holt92a}.  The first QBZ is spanned in the range
$-\hbar\omega/2\leq {\mathcal E}\leq\hbar\omega/2$\ .

In what follows we show, initially, quasi-energy spectra as a function of
the electric field intensity.  It is important to notice that in this work
we choose to keep the field frequency constant ($\hbar\omega=0.5$ meV,
corresponding to $\nu\simeq 0.1$ THz), while ``{\it tuning}" the SL
(varying $E_g$, with $V=0.2$ meV fixed).  This reveals, as will be seen,
to be a useful procedure for a clear identification of different dynamic
localization signatures, for the whole range from $E_g /\hbar\omega\ll 1$
to $E_g /\hbar\omega>1$.

In Fig.1, examples of quasi-energy spectra as function of field intensity
are shown.  In all cases the chains are $L=20$ sites long.  Fig.1(a)
presents the limit of $E_g=0$, i.e., a single and isolated mini-band. The
collapses coincide with zeros of the Bessel function,
$J_0(eaF/\hbar\omega)$, ($eaF/\hbar\omega$: $1.2/0.5=2.4$, $2.75/0.5=5.5$, 
$4.32/0.5=8.64$, etc.) 
reproducing a single mini-band analytical result:
${\mathcal E}(k,\omega)=2VJ_0(eaF/\hbar\omega)\cos ka$.  This result
follows the replacement $\hbar k \longrightarrow \hbar k-eA(t)$
\cite{holt93}.  Fig.1.(b) shows the spectra of two mini-bands formed
by the dimerization procedure with $E_g=0.10$ meV, i.e., the
limit $E_g /\hbar\omega\ll 1$, even conserving remnants of 
the collapses at zeros of $J_0(eaF/\hbar\omega)$, 
analytically described, 
considering the same momentum substitution above\cite{zhao97}, by:

\begin{equation} 
{\mathcal E}(k,\omega)=\pm 2[(E_g)^2+V^2J_0^2({eaF\over
\hbar\omega})\cos ^2({ka\over 2})]^{1/2}. \label{joz} 
\end{equation}

But, a dimerized SL with $E_g/\hbar\omega\sim 1$ shows a qualitative 
different behaviour: first, each mini-band collapses at values of $F$ 
near to corresponding first zero of $J_0(e2aF/\hbar\omega)$, 
since now the SL period doubles, this first
collapse should occur at $eaF\approx 0.6$ meV, and this is already noticeable
for $E_g \simeq\hbar\omega$, Fig.1(c), although evidences of collapsing
mini-bands are observed for $eaF<0.6$ meV. And second,
the dressed mini-bands shown in Fig.1(c) from
top to bottom, are the $m=0$ replica of the $2^{nd}$ mini-band 
and $m=+1$ of the $1^{st}$ mini-band. The AC Stark shifts and the 
coupling of these replicas competes leading 
to anti-crossings at the QBZ edges at $eaF\simeq 2$ meV, 3.5 meV, etc.
representing a field induced one-photon resonance \cite{holt94}.

A clear isolated behaviour of the mini-bands, similar to Fig.1(a), 
occurs if $E_g/\hbar\omega>1$, as can be seen for $E_g/\hbar\omega=3$, and
at low field intensities in Fig.1(d), in which the dressed mini-bands 
depicted from top to bottom are $m=-1$ replica of the $2^{nd}$ mini-band 
and $m=+2$ replica of the $1^{st}$ mini-band.  
We observe a clear mini-band collapse at the first zero of 
$J_0(e2aF/\hbar\omega)$ and a strong anti-crossing - mini-band breakdown - 
at $eaF\simeq 1.5$ meV. For high field intensities one can only see 
mini-band collapses associated to three-photon resonances between mini-band 
replicas.

\begin{figure*}[h] 
\leftline{\includegraphics[scale=0.48]{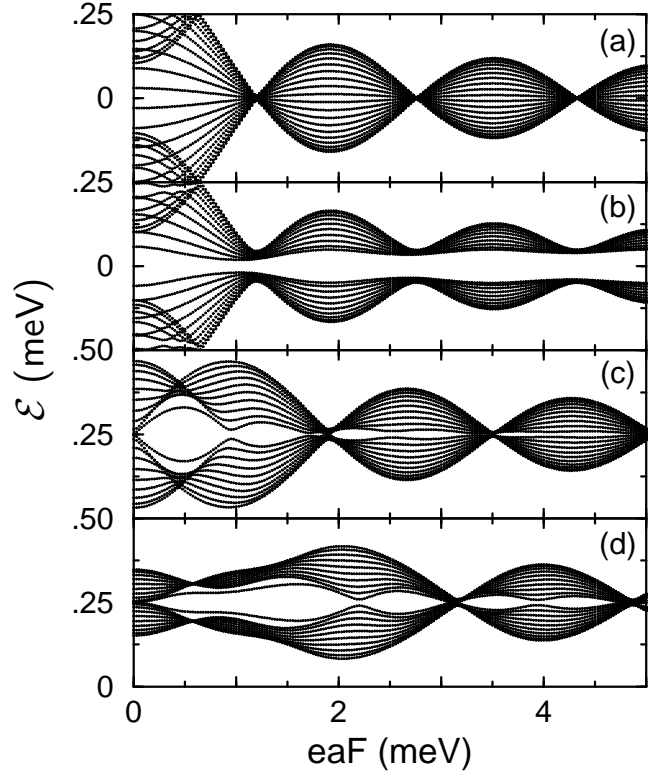}}
\begin{minipage}{8cm}
\caption{Quasi-energy spectra as a function of field intensity for
different chains $L=20$ sites long.  Hopping parameters and field
frequency are the same in all cases, $V=0.2$ meV and $\hbar\omega=0.5$
meV, respectively.  Atomic energies are varied, in order to increase the
mini-band-gap $E_g$ from top to bottom:  (a) $E_g=0$ (single band limit); 
(b)$E_g/\hbar\omega=0.1\ \mbox{meV}/0.5\ \mbox{meV}=0.2$; 
(c) $E_g/\hbar\omega=0.5\ \mbox{meV}/0.5\ \mbox{meV}=1.0$; and 
(d) $E_g/\hbar\omega=1.5\ \mbox{meV}/0.5\ \mbox{meV}=3.0$.  \label{1}}
\end{minipage}
\end{figure*}

Some light may be
shed on the problem by comparing mini-band spectra to the spectra of the
corresponding single dimer unit of the SL. In Fig.2(a), the
mini-band spectra of Fig.1(d), $E_g/\hbar\omega=3$, (small dots), are
compared to the spectra of the two level system corresponding to one dimer
(large dots) that constitutes the SL.  In Fig.2(b) the same is shown for
$E_g/\hbar\omega=3.2$ (out of three-photon resonances).  
Having in mind the dimer spectra, we see that
replicas of these levels cross or anti-cross at the same intensities as
the mini-bands anti-cross. Perfect crossing occur only for the two level
system when $E_g/\hbar\omega=n$, according to the von Neumann-Wigner rule
\cite{holt92a}. Observing the SL and isolated dimer spectra, after the
breakdown of the mini-bands ($eaF\approx 1.5$ meV), 
a new dynamic localization mechanism shows up due to the dominant 
behaviour of the dimers that constitute the SL and not to a field effect 
on the bulk as implicit
in analytical results, like Eq.(\ref{joz}). It is interesting to notice
that this dominant dimer effect signature alternates with field
induced mini-band gap collapses.  Both effects occur only for $E_g
/\hbar\omega=n$ ($n$-photon resonances).  
Indeed, the crossing of the dimer level replicas occur
at zeroes of $J_n(eaF/\hbar\omega)$. The apparent crossings of the 
mini-bands at the QBZ edges are actually anti-crossings not
resolved in the figure.  Detuning the $E_g/\hbar\omega=n$ condition,
Fig.2(b), the mini-band gap collapses disappear and the dynamic
localization due to the dominant behaviour of the dimers 
evolves in a double collapse
structure. This last result, Fig.2(b), was previously observed in Fig.6 of 
reference \cite{hone93}, where the authors take into account the SL
potencial profile explicitly. Resuming, 
these results lead to the following picture: the two
mini-bands behave as isolated ones, only in the limit $E_g /\hbar\omega>1$
and below the mini-band electric breakdown range.  At higher field
intensities dynamical localization manifests as a decoupling
of the SL in the constituting dimers, leading to multi-photon 
resonances between dynamically localized states.  
A related effect are the mini-band gap collapses at certain
field intensities, for which the mini-band dispersions are maxima, when
$E_g/\hbar\omega=n$.

\begin{figure*}[h]
\leftline{\includegraphics[scale=0.48]{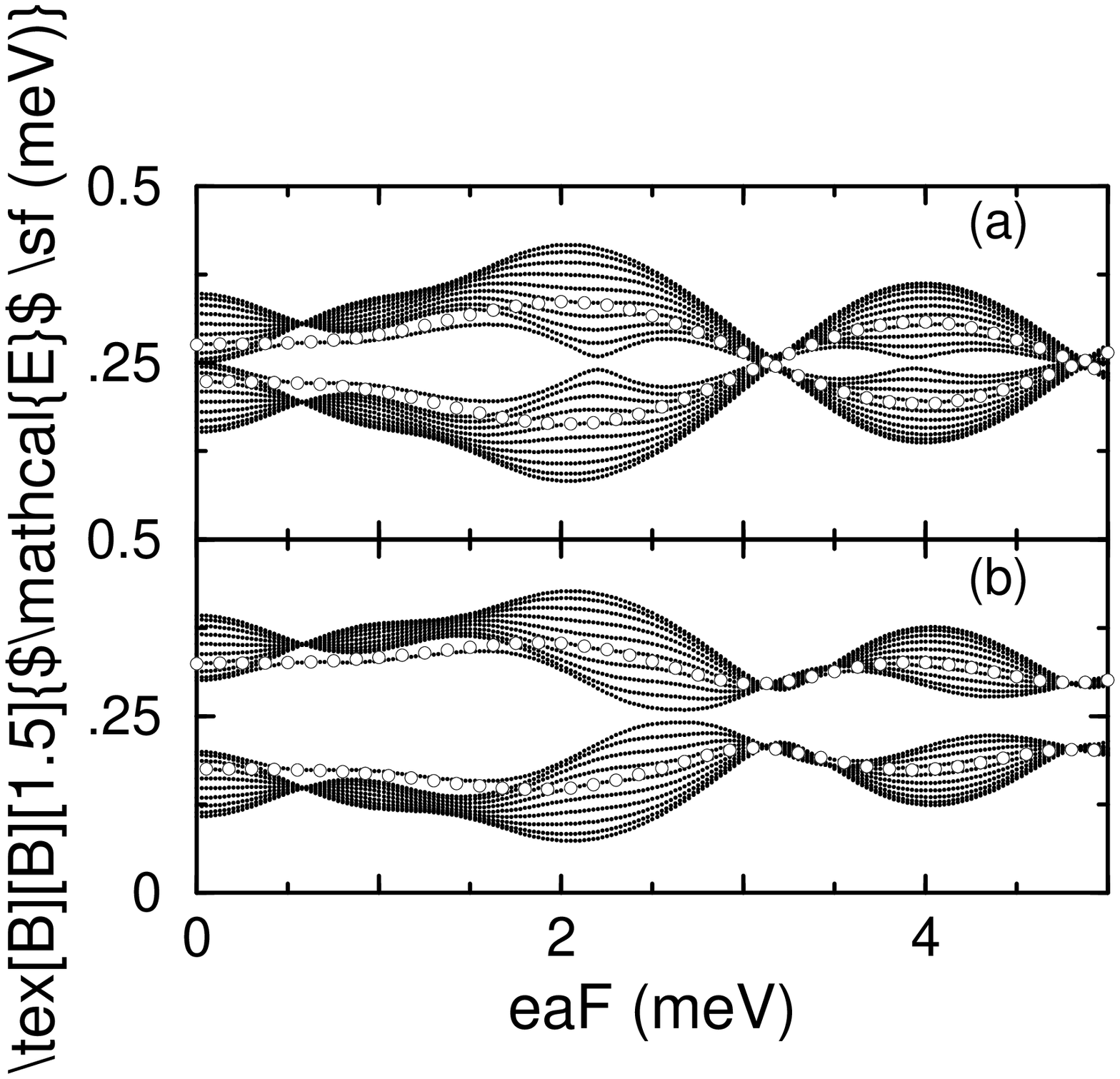}}
\begin{minipage}{8cm}
\caption{Quasi-energies as a function of field intensity, comparing finite
chains (small dots), $L=20$ sites long, with corresponding dimers, two
level systems (large dots).  
(a) $E_g/\hbar\omega=1.5\ \mbox{meV}/0.5\ \mbox{meV}=3.0$; and
(b) $E_g/\hbar\omega=1.6\ \mbox{meV}/0.5\ \mbox{meV}=3.2$.  \label{2}} 
\end{minipage}
\end{figure*}

This picture for the evolution of dressed mini-bands, is clearly 
illustrated by analyzing the projection of mini-band width $\Delta$,
mapped on a $E_g$ {\it versus} $eaF$ plane shown in
Fig.3, with $\hbar\omega=0.5$ meV fixed, while we are ``{\it tuning}"
the SL mini-band gap, considering a fine mesh for varying $E_g$. 
In the brightness scale, dark is for
very small mini-band width, indicating mini-band collapses, and white 
is for large mini-band width, i.e., field intensity versus $E_g$ regions of
relative maximum mini-band broadening. 

Near $E_g=0$ we see the dark spots at $eaF$ values corresponding
to the single band collapse of Fig.1(a).  For $E_g>1$ meV collapsing of
individual mini-bands of the dimerized SL appear at zeroes of
$J_0(e2aF/\hbar\omega)$, indicated by grey vertical lines.  These
collapses appear successively with increasing $E_g$ in the range
$E_g>eaF$.  Indeed, slightly below the $E_g = eaF$ diagonal a bright region
indicates the dynamic breakdown regime, separating collapses of
qualitative different evolution in the quasi energy spacing map. For
$E_g < eaF$ the positions of mini-band collapses are functions of the field
intensity: these are collapses related to the dominant behaviour of  
the dimers. The
single spots at $E_g/\hbar\omega=n$ are for crossings in the two level
(dimer) systems.  The two dark lines connecting single spots are the
double collapsing structure that evolves out of the resonance condition,
like in Fig.2(b).  The horizontal grey lines at the left reflect an
approximately constant degeneracy of surface states, for field intensities
up to the dynamical breakdown, when mini-band replica strongly overlap in
this representation. In this map, the upper left and lower
right half-planes represent qualitatively different physical situations;
respectively the isolated mini-band regime and a regime of field induced
rearrangement of the electronic structure.

\begin{figure*}
\psfrag{e}[B][B][1.5]{\sf eaF (meV)}
\psfrag{E}[B][][1.5]{$\mathsf{E_g}$\sf (meV)} 
\rightline{\includegraphics[scale=0.50]{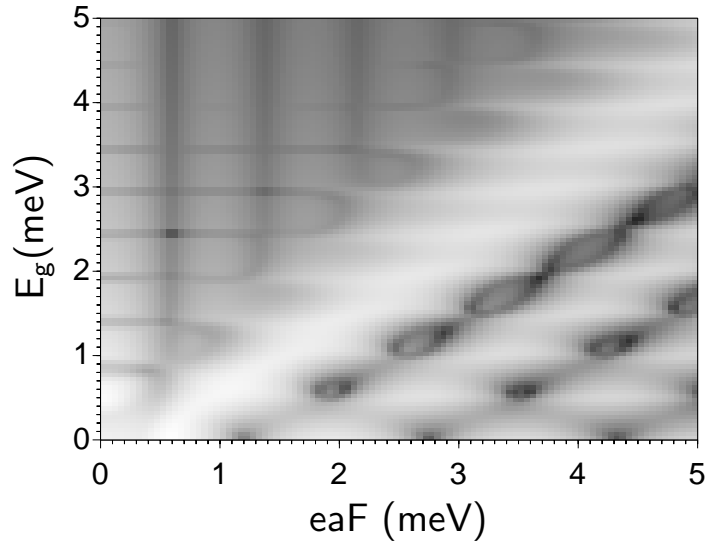}}
\begin{minipage}{8cm}
\caption{Projection of the mini-band width $\Delta$ on a $E_g$ {\sl versus}
$eaF$ representation plane.  The brightness scale spans from black (small
$\Delta$:  collapsing mini-bands) to white (widest $\Delta$: large
mini-band dispersion).  \label{3}} 
\end{minipage}
\end{figure*}

This mapping refers to the mini-band dispersion, indicating
clearly the evolution of mini-band collapses from one dynamic 
localization regime into another as a function of field intensity. 
However, there are also important mini-band gap variations, which 
apparently approaches zero in
the main breakdown region, as well as in the ``satellite" breakdown
regions, when $E_g/\hbar\omega=n$, as can be seen in the case of Fig.2(a).
Referring to Fig.2(a), one sees that at the intensities for which the 
mini-band gaps seem to collapse, the spectra resembles a single band 
density of states. For this reason,
quasi-energy spectra as densities of states for selected field intensities
for the case shown in Fig.2(a) are depicted in Fig.4. We
clearly see a transition from a two mini-band density of states, Fig.4(a),
(low field intensity range) to a single band density of
states, Fig.4(c).  Increasing further the field
intensity, the two mini-band situation can be recovered, Fig.4(d).  The
shape of the density of states, shows an effective electronic structure
alternating from binary (Peierls instability) to single (suppression of 
Peierls instability) band with field intensity. Such
effect is an AC field tunable Peierls instability. 
This opens the possibility of a new 
insulator-metal transition, considering dimerized SLs doped in such a way 
that the first bare mini-band is fully occupied. Since it has
been shown that consequences of band collapses in a single-band
tight-binding model survives in the presence of Coulomb interactions
\cite{koch}, one could expect to observe insulator-metal like transitions
in such dimerized SLs driven by tunable intense fields in the THz range.

\begin{figure*} 
\centerline{\includegraphics[scale=0.5]{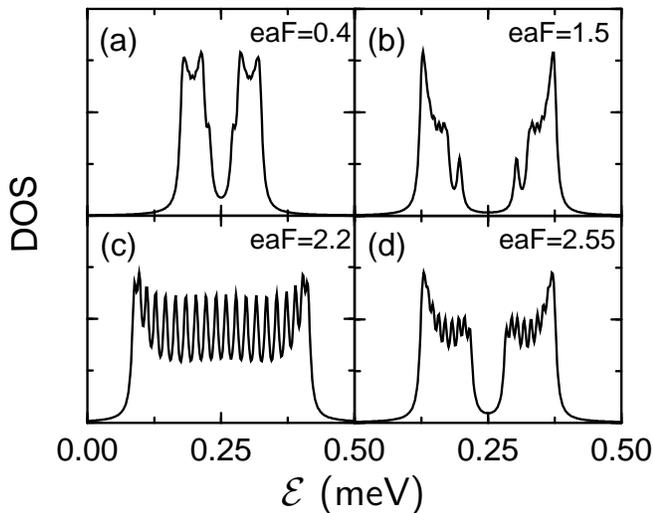}}
\begin{minipage}{8cm}
\caption{Density of states obtained from quasi-energy spectra at given
field intensities, using a Breit-Wigner fit, 
for the case depicted in Fig.1(d).  (a) $eaF=0.3$ meV,
(b) $eaF=1.5$ meV, (c) $eaF=2.2$ meV, and (d) $eaF=2.55$ meV. 
\label{4}}
\end{minipage}
\end{figure*}

Other results, based on a different heuristic 
model \cite{jauho95,drese96}, do not show a clear dynamic breakdown 
and signatures of isolated mini-bands behaviour survive up to very 
intense fields. These results consider also alternation in barrier 
thicknesses and, therefore, do not necessarily contradict the present 
ones. Including alternation of hopping parameters to our model, a similar 
picture to these previously reported results is obtained. 
The comparison of the present results with some previous reported ones 
suggests that further work is still necessary in order to build up a 
systematic understanding of the behaviour of dressed coupled mini-bands.
 
In summary, we showed the existence of two dynamic localization regimes in
SLs dimerized by alternating well widths. No memory of the low field regime 
survives beyond the dynamic breakdown region, where a insulator-metal
transition may occur as a consequence of the suppression of a Peierls-like 
instability for resonance field frequencies.

The authors acknowledge CAPES, FAPESP and CNPq for financial support.
P.A.S. is grateful to P. Hawrylak for introducing him to the subject of
this work, and P. H. R. thanks to M. Wagner for critical reading of 
the manuscript.

\end{multicols}

\end{document}